\documentclass[conference]{IEEEtran}
\usepackage{cite}
\usepackage{amsmath,amssymb,amsfonts}
\usepackage{algorithmic}
\usepackage{graphicx}
\usepackage{textcomp}
\usepackage{xcolor}
\usepackage{float}
\usepackage{hyperref}
\usepackage{booktabs}
\hypersetup{colorlinks=true, 
breaklinks=true, 
urlcolor= black, 
linkcolor= black,
citecolor= black 
} 

\usepackage{cleveref} 
\Crefname{figure}{Fig.}{}
\Crefname{equation}{}{}
\Crefname{paragraph}{paragraph}{}

\usepackage{braket}
\newcommand\ketbra[1]{\ensuremath{ \ket{#1} \bra{#1} }}

\usepackage{scalerel,stackengine}
\stackMath
\newcommand\reallywidehat[1]{%
\savestack{\tmpbox}{\stretchto{%
  \scaleto{%
    \scalerel*[\widthof{\ensuremath{#1}}]{\kern-.6pt\bigwedge\kern-.6pt}%
    {\rule[-\textheight/2]{1ex}{\textheight}}
  }{\textheight}%
}{0.5ex}}%
\stackon[1pt]{#1}{\tmpbox}%
}

\usepackage{acro}
\DeclareAcronym{vqa}{short=\textsc{VQA}, long=Variational Quantum Algorithm}
\DeclareAcronym{vqe}{short=\textsc{VQE}, long=Variational Quantum Eigensolver}
\DeclareAcronym{adapt}{short=\textsc{ADAPT}, long=Adaptive Derivative-Assembled Pseudo-Trotter}
\DeclareAcronym{vqls}{short=\textsc{VQLS}, long=Variational Quantum Linear system Solver}
\DeclareAcronym{qpu}{short=\textsc{QPU}, long=Quantum Processing Unit}
\DeclareAcronym{qaoa}{short=\textsc{QAOA}, long=Quantum Approximate Optimization Algorithm}
\DeclareAcronym{aqc}{short=\textsc{AQC}, long=Adiabatic Quantum Computing}
\DeclareAcronym{isq}{short=\textsc{ISQ}, long=Intermediate Scale Quantum}
\DeclareAcronym{nisq}{short=\textsc{NISQ}, long=Noisy Intermediate Scale Quantum}
\DeclareAcronym{ftqc}{short=\textsc{FTQC}, long=Fault Tolerant Quantum Computing}
\DeclareAcronym{hva}{short=\textsc{HVA}, long=Hamiltonian Variational Ansatz}
\DeclareAcronym{uccva}{short=\textsc{UCCVA}, long=Unitary Coupled Cluster Variational Ansatz}
\DeclareAcronym{heva}{short=\textsc{HEVA}, long=Hardware Efficient Variational Ansatz}
\DeclareAcronym{tva}{short=\textsc{TVA}, long=Truncated Variational Ansatz}
\DeclareAcronym{pva}{short=\textsc{PVA}, long=Projector Variational Ansatz}
\DeclareAcronym{qsp}{short=\textsc{QSP}, long=Quantum Signal Processing}
\DeclareAcronym{aa}{short=\textsc{AA}, long=Amplitude Amplification}
\DeclareAcronym{qpe}{short=\textsc{QPE}, long=Quantum Phase Estimation}
\DeclareAcronym{scb}{short=\textsc{SCB}, long=Single Component Basis}
\DeclareAcronym{hf}{short=\textsc{HF}, long=Hartree-Fock}
\DeclareAcronym{uccsd}{short=\textsc{UCCSD}, long=Unitary Coupled-Cluster Singles and Doubles}
\DeclareAcronym{lcu}{short=\textsc{LCU}, long=Linear Combination of Unitaries}
\DeclareAcronym{hc}{short=\textsc{HC}, long=Hermitian Conjugate}

\usepackage[colorinlistoftodos, textsize=small]{todonotes}

\usepackage{tikz}
\usetikzlibrary{positioning,arrows,calc,math,angles,quotes}
\usepackage{multirow}

\def\BibTeX{{\rm B\kern-.05em{\sc i\kern-.025em b}\kern-.08em
    T\kern-.1667em\lower.7ex\hbox{E}\kern-.125emX}}
    
\begin{document}


\title{
Projector Quantum Variational Ansatz
}

\def\orgadep{CEA, List, F-91120}
\def\orga{CEA}
\def\loc{Palaiseau, France} 
\def\univ{Université Paris-Saclay}

\author{\IEEEauthorblockN{Thomas DUMONTIER}
\IEEEauthorblockA{\textit{\univ} \\
\textit{\orgadep}\\
\loc \\
0009-0005-6388-6288}
\and
\IEEEauthorblockN{Robin OLLIVE}
\IEEEauthorblockA{\textit{\univ} \\
\textit{\orgadep}\\
\loc \\
0009-0006-7539-363X} 
\and
\IEEEauthorblockN{Stephane LOUISE}
\IEEEauthorblockA{\textit{\univ} \\
\textit{\orgadep}\\
\loc \\
0000-0003-4604-6453}
}

\maketitle

\begin{abstract}
  
  Quantum computing offers several algorithms for computing the ground state of a problem Hamiltonian.
  Algorithms belonging to the \ac{ftqc} regime, such as \ac{qpe} and \ac{qsp}, are the most desirable because they offer optimal asymptotic scaling.
  In the \ac{nisq} regime, the most realistic approaches involve \ac{vqe} algorithms and their variants.
  A central difference between the two regimes is that \ac{ftqc} algorithms do not prepare the ground state via direct state transition. Instead, they build a projector that identifies the ground state using ancillary qubits that flag the target subspace. 
  The desired state is then recovered via amplitude amplification or post-selection. 
  
  
  In this work, we propose a \ac{vqe} ansatz whose structure is more similar to that of an \ac{ftqc} algorithm. We introduce the \ac{pva}, which imports the projector structure into the adaptive variational setting.
  Depending on its parametrization, this ansatz can be equivalent to either an \ac{adapt}-\ac{vqe} when all ancilla mixing angles vanish or to an \ac{isq}-\ac{qsp} quantum circuit structure.
  
  We benchmarked \ac{pva} against an \ac{adapt}-\ac{vqe} baseline using both qubit-excitation and fermionic operator pools.
  We propose an implementation in which an $L$-layer \ac{pva} circuit is designed using $ 3L+1$ parameters and requires only one ancilla and two additional CNOT gates per Pauli generator used in \ac{adapt}.
  Our experimental results show that this first proposal of \ac{pva} converges using a smaller number of CNOTs gates than the standard \ac{adapt}-\ac{vqe} on the system tested.

\end{abstract}


\acresetall
\section{Introduction}
\acp{vqa} constitute a class of continuous optimization algorithms \cite{cerezo_variational_2021, barzen_patterns_2021}. 
In \acp{vqa}, the \ac{qpu} is used to evaluate a parametrized quantum circuit called ansatz.
Then, a classical optimizer updates the circuit parameters in order to minimize a cost function.
The ansatz is designed to be sufficiently expressive to represent the target state while remaining shallow enough for \ac{nisq} hardware. Ideally, the search space is large enough to reach the minimum of the cost function, but not so large that the optimizer encounters a barren plateau.
The first \ac{vqa} proposal is the \ac{vqe} algorithm \cite{peruzzo_variational_2014, yung_transistor_2014, mcclean_theory_2016, fedorov_vqe_2022} which finds ground states variationally.
Extensions of this algorithm have since been developed to access excited states \cite{peruzzo_variational_2014,zhang_variational_2020}.
The VQE cost function aims to minimize the expectation value of the problem Hamiltonian, obtained via projective measurement, evaluated on an optimized state vector.

A central challenge in VQAs is constructing an ansatz that balances expressivity, circuit depth, and gate count. The ansatz needs to be as shallow as possible in order to run on NISQ devices.
Proposing new strategies for constructing an ansatz is a very active field since the proposal of the first \ac{vqa}.
It is possible to classify most of the ansatzes as variants of a few ansatz generation strategies.
\ac{heva} \cite{kandala_hardware-efficient_2017}.
are built from specific \ac{qpu}-native gates.
Problem-inspired ansatzes that are inspired by other strategies used to tackle the same problem.
Notable examples include the \ac{uccva} \cite{peruzzo_variational_2014, yung_transistor_2014, shen_quantum_2017, mcclean_theory_2016} inspired by classical strategies to address chemistry problems and their variants \cite{babbush_low-depth_2018, barkoutsos_quantum_2018, gard_efficient_2020, herasymenko_diagrammatic_2021, choquette_quantum-optimal-control-inspired_2021}.
This category of ansatzes also encompasses the \ac{hva} which is inspired by the \ac{aqc} principles \cite{farhi_quantum_2000} and all its variants \cite{wecker_progress_2015, cade_strategies_2020, reiner_finding_2019, park_hamiltonian_2024, wiersema_exploring_2020, ho_efficient_2019}.
When applied to an optimization problem, this \ac{hva} leads to the \ac{qaoa}, which also has many variants \cite{farhi_quantum_2014, farhi_quantum_2019, blekos_review_2024, hadfield_quantum_2019, yoshioka_fermionic_2023}. 
The different types of ansatz are classified in \Cref{table_ansatz_zoo}.

\begin{table*}[tb]
  \caption{Ansatz basic classification. \\
  \textmd{\scriptsize Reduced search-space with convergence guarantee for sufficiently large ansatz corresponds to the label 'constrained search-space'.
  If the variational circuit can be constructed only with the cost-function knowledge, it is referred to by the label 'Any Problem'.}
  }
  \center
    {\begin{tabular}{|ccc|c|c|c|c|c|c|}
		\hline
		\multicolumn{3}{|c|}{\multirow{2}{*}{\textbf{Ansatz origin}}} & \multirow{2}{*}{\textbf{Short}} & \multicolumn{2}{|c|}{\textbf{Problem (by Design)}} & \textbf{Layer-by-layer} & \multirow{2}{*}{\textbf{Constrained}} & \multirow{2}{*}{\textbf{Any}} \\
		\cline{5-6}
		\multicolumn{3}{|c|}{\multirow{2}{*}{Truncated and parametrized algorithm}} & \multirow{2}{*}{\textbf{Name}} & eigenvalue & linear & \textbf{building} & \multirow{2}{*}{\textbf{Search-Space}} & \multirow{2}{*}{\textbf{Matrix}} \\
		\multicolumn{3}{|c|}{} & & research & system & (ADAPT) & & \\
		\hline
		\multirow{2}{*}{Physically inspired} & \multirow{2}{*}{$\longrightarrow$} & \multirow{2}{*}{UCC(SDT...)} & \multirow{2}{*}{\acs{uccva}} & \multirow{2}{*}{\text{\color{green} \rlap{$\checkmark$}}} & \multirow{2}{*}{\color{red} $\times$} & \multirow{2}{*}{\text{\color{green} \rlap{$\checkmark$}}} & \multirow{2}{*}{\text{\color{green} \rlap{$\checkmark$}}} & \multirow{2}{*}{\color{red} $\times$} \\
		 & & & & & & & & \\
		\hline
		\multirow{2}{*}{Adiabatic evolution} & \multirow{2}{*}{$\longrightarrow$} & Variational & \multirow{2}{*}{\acs{hva}/\ac{qaoa}} & \multirow{2}{*}{\text{\color{green} \rlap{$\checkmark$}}} & \multirow{2}{*}{\color{red} $\times$} &  \multirow{2}{*}{\text{\color{green} \rlap{$\checkmark$}}} & \multirow{2}{*}{\text{\color{green} \rlap{$\checkmark$}}} & \multirow{2}{*}{\text{\color{green} \rlap{$\checkmark$}}} \\
		 & & Hamiltonian &  & & & & & \\
		\hline
		Eigenvalue polynomial & \multirow{2}{*}{$\longrightarrow$} & \multirow{2}{*}{QSP based} & \multirow{2}{*}{\acs{pva}} & \multirow{2}{*}{\text{\color{green} \rlap{$\checkmark$}}} & \multirow{2}{*}{\text{\color{green} \rlap{$\checkmark$}}} &  \multirow{2}{*}{\text{\color{green} \rlap{$\checkmark$}}} & \multirow{2}{*}{\text{\color{green} \rlap{$\checkmark$}}} & \multirow{2}{*}{\text{\color{green} \rlap{$\checkmark$}}} \\
		transformation ({\color{red} our work}) & & & & & & & & \\
		\hline
		\hline
		\multicolumn{3}{|c|}{Hardware Efficient} & \acs{heva} & \text{\color{green} \rlap{$\checkmark$}} & \text{\color{green} \rlap{$\checkmark$}} &  & {\color{red} $\times$} & \text{\color{green} \rlap{$\checkmark$}} \\
		\hline
	\end{tabular}}
  \label{table_ansatz_zoo}
\end{table*}

All the previously mentioned ansatzes have a static quantum circuit.
Another approach proposed in \ac{adapt}-\ac{vqe} \cite{grimsley_adaptive_2019} consists of assembling the circuit iteratively, depending on the gradient associated with the different parameterized blocks that can be added. 
This algorithm also has variants that often differ in the pool of operators that can be added and in how blocks are appended to the quantum circuit. 
For example, the original ADAPT-VQE utilizes a physically inspired pool of fermionic excitation operators, while Qubit-ADAPT-VQE \cite{tang_qubit-adapt-vqe_2021, feniou_overlap-adapt-vqe_2023, anastasiou_how_2023, viswanathan_optimal_2025} uses individual Pauli strings in order to reduce the two-qubit gate depth per iteration. On the other hand, CEO-ADAPT-VQE \cite{ramoa2025reducing} introduces a ``Coupled Exchange Operator'' pool designed to reduce measurement overhead and improve hardware efficiency. We can also mention Tetris-ADAPT-VQE \cite{anastasiou2024tetris} that evaluates commutativity to add multiple non-overlapping operators simultaneously, and Geo-ADAPT-VQE \cite{sohail2026geo}, which uses a selection rule based on natural gradient to grow the ansatz along the true geometry of the quantum state space.
Finally, it is interesting to mention that \ac{adapt}-\ac{vqe} also outperforms usual \ac{vqe} for the Fermi-Hubbard model, an instance of solid-state physics \cite{gyawali_adaptive_2022}.

Other \ac{vqa} based on more complex cost functions have emerged \cite{cerezo_variational_2021}.
Popular examples include Variance optimization for excited states \cite{zhang_variational_2020} and \ac{vqls} for matrix inversion. 
The ansatzes proposed for these different algorithms are mostly hardware-efficient ansatz or \ac{vqe} ansatz  \cite{bravo-prieto_variational_2020,xu_variational_2021,pellow-jarman_near_2023,rao_performance_2024,gnanasekaran_efficient_2024,marfany_identifying_2024}. 

Iterative (\textit{i.e.} non-variational) quantum algorithms allow for searching for the smallest eigenvalue of Hermitian matrices.
Two popular algorithms of this category are \ac{qpe} \cite{kitaev_quantum_1995}, and \ac{qsp} \cite{martyn_grand_2021} which have an \ac{isq} variant \cite{dong_ground_2022, ollive_quantum_2024, ronfaut_numerical_2026}.
Both algorithms use at least one ancilla qubit to indicate whether the state entangled with the other register is the ground state, in other words, it behaves as a projector-controlled-NOT gate \cite[Section C]{martyn_grand_2021}.

Our work proposes a new method for constructing an ansatz inspired by the \ac{ftqc} algorithm: the \ac{pva}.
The remainder of the paper can be decomposed as follows.
\Cref{sec_ansatz_construct} contains the theoretical aspects of the ansatz construction. 
The general procedure is described and then applied to construct a new ansatz. 
This ansatz, inspired by the \ac{qsp} algorithm, defines a new family of algorithms that includes both standard \ac{adapt}-\ac{vqe} and \ac{qsp}-like \ac{isq} circuits.
\Cref{sec_expe_set} describes the experimental choices done to evaluate our ansatz.
\Cref{sec_result} proposes an experimental comparison of these ansatzes' performance with respect to the standard \ac{adapt}-\ac{vqe}. 

\section{Ansatz Construction \label{sec_ansatz_construct}}
\subsection{Common Requirement for ADAPT-VQE and ISQ-QSP}
\paragraph{Problem Hamiltonian} The Hamiltonians we try to diagonalize are described thanks to a weighted sum of Hermitian matrices:
\begin{equation}
  \widehat{H_{p}} = \sum_{i} \alpha_{i} \widehat{H}_{i}
\end{equation}
with all the $\alpha_{i}$ are real numbers and each $\widehat{H}_{i}$ matrices that can be both evaluated and Hamiltonian simulated on a quantum computer with an easy basic gate decomposition.
This decomposition serves two purposes in the context of this work:
\begin{itemize}
    \item The ability to measure the expectation value of a vector on this Hamiltonian to construct the cost function:
    \begin{equation}
        \bra{\psi} \widehat{H_{p}} \ket{\psi} = \sum_{i = 1} \alpha_{i} \bra{\psi} \widehat{H}_{i} \ket{\psi}
    \end{equation}
    with $ \widehat{H_{i}} = \widehat{V_{i}}^{\dag} \widehat{D} \widehat{V_{i}} $, $\widehat{D}$ a diagonal matrix and $\widehat{V_{i}}$ a unitary matrix.
    \item The possibility to approximate the problem query by Trotterization (\ac{qsp}) or to compose the pool (\ac{adapt}-\ac{vqe}) by a combination of the Hamiltonian simulation of these terms:
    \begin{equation}
        e^{i t \widehat{H_{p}}} \simeq \prod_{i = 1} e^{i t \alpha_{i} \widehat{H_{i}}}
        \label{eq_trotterization}
    \end{equation}
\end{itemize}
Typically, a chemistry problem can be decomposed either using Pauli mapping or directly with the chemistry-native ladder and the number of excitation operators \cite{ollive_gate_2025}.

\subsection{\acl{isq}-\acl{qsp}}
Construction of the \acl{qsp} based Ansatz is motivated because the \ac{qsp} algorithm\cite{martyn_grand_2021} uses only one extra ancilla qubit to process the query.
It is possible to construct the query using only the same ancilla qubit as \ac{qsp}.
The query is approximated by Trotterization \cite{childs_theory_2021, ronfaut_numerical_2026}.
This idea of using \ac{qsp} based on the Hamiltonian simulation of the problem Hamiltonian $\widehat{H_{p}}$ 
is exploited in \cite{dong_ground_2022, ollive_quantum_2024, ronfaut_numerical_2026} to construct \ac{isq} algorithms for eigenstate filtering.
The access to the Hamiltonian simulation of the individual terms of the decomposition $ e^{i t \widehat{H_{i}}} $ is supposed to be known.

\begin{figure*}[tb]
    \begin{center}
        \resizebox{\linewidth}{!}
        {\includegraphics{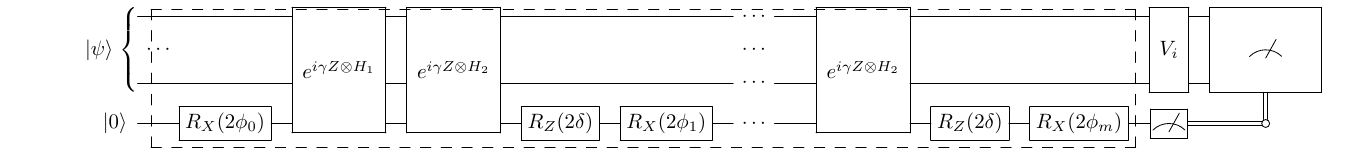}}
    \end{center}
    \caption{Quantum circuit of the \ac{isq}-\ac{qsp} algorithm that measures the expectation value of the summand $\widehat{H_{i}}$.
    Here: $ \widehat{H_{p}} = \widehat{H_{1}} + \widehat{H_{2}} $.
    The dashed box delimits the \ac{qsp} part; the rest of the circuit is the $\widehat{H_{i}}$ expectation value measurement on the lower part of the energy spectrum.
    Note that the $\widehat{H_{p}}$ expectation value is upper bounded by the value $\Delta$ by the filtering done by the \ac{qsp}.}
    \label{circuit_isq_qsp}
\end{figure*}

The following points describe a step-by-step procedure to construct the \ac{isq}-\ac{qsp} projector quantum circuit (illustrated by \Cref{circuit_isq_qsp}):
\begin{enumerate}
  \item Construct the $W_{Z}$ signal operator (the query) associated with a Hamiltonian simulation using a product formula:
  \begin{equation}
  \begin{aligned}
    \widehat{W_{Z}}[\widehat{H_{p}}] & = 
        \begin{bmatrix}
            e^{i (\gamma \widehat{H_{p}} + \delta \widehat{I})} & 0 \\
            0 & e^{- i (\gamma \widehat{H_{p}} + \delta \widehat{I})}
        \end{bmatrix}
        \begin{matrix}
          \ket{\psi} \ket{0} \\
          \ket{\psi} \ket{1}
        \end{matrix} \\
        & = \ketbra{\lambda_{i}} \otimes
        \begin{bmatrix}
            e^{i (\gamma \lambda_{i} + \delta)} & 0 \\
            0 & e^{- i (\gamma \lambda_{i} + \delta)}
        \end{bmatrix}
        \begin{matrix}
          \ket{0} \\
          \ket{1}
        \end{matrix} \\
        & = e^{i \widehat{Z} \otimes (\gamma \widehat{H_{p}} + \delta \widehat{I})} \\
        & \sim e^{i \delta \widehat{Z} \otimes \widehat{I}} \prod_{i} e^{i \gamma \alpha_{i} \widehat{Z} \otimes \widehat{H_{i}}} \\
  \end{aligned}
  \end{equation}
  With $\widehat{H_{p}} = \sum_{i} \lambda_{i} \ketbra{\lambda_{i}},  \gamma = \frac{\pi}{2} \frac{\Gamma}{|\widehat{H_{p}}|}$ with $\Gamma \leq 1$
  \begin{itemize}
    \item The Hamiltonian normalization factor is a least $ \gamma \leq \frac{\pi}{2} \frac{1}{|\widehat{H_{p}|}} $ so that the eigenphases are defined lies in $ [ - \Gamma \frac{\pi}{2} ; \Gamma \frac{\pi}{2} ] $.
    With $\Gamma = 1$ and a boxcar function $\mathrm{f}$, shifting $\delta$ over $ [ 0; \pi ] $, allows to project all the eigenvalues in the two parts of the block-encoding to separate them by their norm: the one under and the one bellow $\Delta$ with $ \delta \pm \frac{\pi}{2} = \gamma \Delta $.
  \end{itemize}
  \item Construct a \ac{qsp} algorithm:
  \begin{equation}
    \widehat{QSP}_{Z}[\mathrm{f}] = \prod_{m = 1}^{m_{\phi}} \{ \widehat{S_{X}}(\phi_{m}) \widehat{W_{Z}} \} \widehat{S_{X}}(\phi_{0})
  \end{equation}
  with $\underline{\phi}$ the phase angles associated with the function $\mathrm{f}$ of interest, $m_{\phi}$ the number of phase angles and the signal processors:
  \begin{equation}
  \begin{aligned}
    \widehat{S_{X}}(\theta) & = e^{i \theta \widehat{X}} = \widehat{R_{X}}(2 \theta)
  \end{aligned}
  \end{equation}
  \begin{itemize}
    \item The phase angles are completely defined by the choice of the function.
    The bigger the number of phase angles, the better the function approximation.
    To construct a polynomial approximation of a projector, a relevant function choice is a boxcar function\footnote{
        To solve \ac{vqls}, one can change the value of the shift and use a phase angle series associated with the inverse function.
    } such that the ancilla qubit indicates a projection onto the lower part of the spectrum:
    \begin{equation}
        \bra{0} \widehat{QSP}_{Z}[\mathrm{f}, \widehat{H_{p}}, \delta] \ket{0} \ket{\psi} = \sum_{i, \lambda_{i} < \Delta} \ketbra{\lambda_{i}} \ket{\psi}
    \end{equation}
  \end{itemize}
\end{enumerate}

\subsection{ADAPT-\acl{vqe}}
The \ac{adapt}-\ac{vqe} \cite{grimsley_adaptive_2019} requirement to generate an ansatz is a pool of operators that are:
\begin{itemize}
    \item easy to measure the expectation value on the \ac{qpu} state.
    \item whose Hamiltonian simulation can be implemented on the \ac{qpu}.
\end{itemize}
The pool selected for our work contains all the summands that compose the problem Hamiltonian to diagonalize  $ \mathbb{P} =  \{ \widehat{H_{i}} | \widehat{H_{p}} = \sum_{i} \alpha_{i} \widehat{H}_{i} \} $.
These summands are either Pauli-string $ \widehat{H_{i}} \in \widehat{PS} $, the tensorial product of Pauli matrices, or the fermionic ladder operator mapped using the \ac{scb} formalism \cite{ollive_gate_2025}.

\begin{figure}[tb]
    \begin{center}
        {\includegraphics{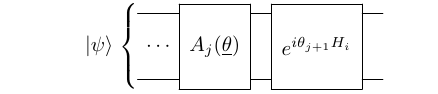}}
    \end{center}
    \caption{Quantum circuit of the \ac{adapt}-\ac{vqe} ansatz at the $ j + 1 $ iteration.}
    \label{circuit_dapt_vqe}
\end{figure}

\ac{adapt}-\ac{vqe} contains two nested loops and start in the state $\ket{\psi}$.
The outer loop updates the ansatz by adding the Hamiltonian simulation of the operator from the pool with the largest gradient:
\begin{equation}
    \widehat{A_{j + 1}}(\underline{\theta}) = e^{i \theta_{j + 1}\widehat{H_{i(j)}}}  \widehat{A_{j}}(\underline{\theta})
\end{equation}
with $\widehat{A_{j}}(\underline{\theta})$ the ansatz at the $j^{\text{th}}$ iteration of the outer loop.
For each operator of the pool, this gradient is computed thanks to the evaluation of a commutator expectation value:
\begin{equation}
    \left. \frac{\partial \bra{\psi_{j}} \widehat{H_{p}} \ket{\psi_{j}} }{\partial \theta_{j + 1}} \right|_{\theta_{j + 1} = 0} 
    = i\bra{\psi_{j}} [\widehat{H_{p}}, \widehat{H_{i}}] \ket{\psi_{j}}
    \label{eq_adapt_grad}
\end{equation}
with $ \ket{\psi_{j}} = \widehat{A_{j + 1}}(\underline{\theta}) \ket{\psi} $.
The inner loop optimizes the new ansatz using \ac{vqe}, starting from the angles from the previous optimization, and updates the optimized parameter values.
When the algorithm is run, it generates an ansatz similar to the form shown in \Cref{circuit_dapt_vqe}.

\subsection{Projector Variational Ansatz}
The construction of \ac{pva} is motivated by truncating the signal-processing architecture of \ac{qsp}. Instead of applying the complete Hamiltonian-simulation query at each step, we truncate this product and keep only one conditional evolution. The operators are selected adaptively using the \ac{adapt}-\ac{vqe} framework. Since this truncation removes the analytical \ac{qsp} conditions that determine the evolution times and the signal-processing phases, we have the apparition of $3$ new variational parameters corresponding to these quantities: 
\begin{equation}
    \gamma\alpha_{i_j}\rightarrow\theta_j,
    \qquad
    \delta\rightarrow\delta_j,
    \qquad
    \phi_j^{\mathrm{QSP}}\rightarrow\phi_j.
\end{equation}

This variant of the \ac{adapt}-\ac{vqe} algorithm differs from the original \ac{adapt}-\ac{vqe} by using one ancillary qubit that is used to filter the solutions.
It starts on the state $ \ket{\psi} \ket{0} $.
The selection of the operator of the pool is done with the same commutator measurement but in the subspace in which the eigenstates are filtered with respect to their energy: \begin{equation}
\begin{aligned}
    \bra{\varphi_{j}} [\widehat{H_{p}}, \widehat{H_{i}}] \ket{\varphi_{j}} 
    & \propto \frac{|\bra{\psi_{j}} (\ketbra{0} \otimes [\widehat{H_{p}}, \widehat{H_{i}}]) \ket{\psi_{j}}|}{|\bra{\psi_{j}} (\ketbra{0} \otimes \widehat{I}) \ket{\psi_{j}} |}
\end{aligned}
\end{equation}
with $ \ket{\psi_{j}} = \widehat{A}_{j}(\underline{\theta}, \underline{\delta}, \underline{\phi}) \ket{\psi} = \alpha \ket{\varphi_{j}} \ket{0} + \sqrt{1 - |\alpha|^{2}} \ket{\varphi_{j \perp}} \ket{1} $ with $ |\alpha| \in [0, 1] $.
This is the same gradient used in \ac{adapt}-\ac{vqe} but an approximation to the true gradient of the \ac{pva} cost function with respect to the $\theta_j$ parameters at $(\theta_j, \delta_j, \phi_j)=(0,0,0)$.
We decided to use this gradient because it is the cheapest available criterion and sufficient in practice for \ac{pva}.
We can estimate this gradient using the accepted shots that we already collected for the energy, so no additional circuit executions beyond the commutator itself are required.

\begin{figure}[tb]
    \begin{center}
        \resizebox{\linewidth}{!}
        {\includegraphics{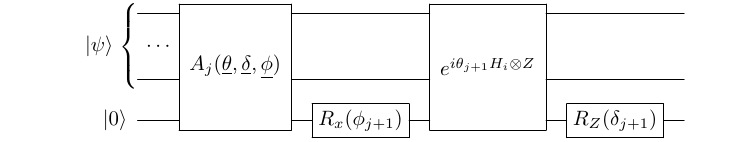}}
    \end{center}
    \caption{Quantum circuit of the Projector-\ac{adapt}-\ac{vqe} ansatz at the $ j + 1 $ iteration.}
    \label{circuit_p_adapt_vqe}
\end{figure}

The operator used to grow the ansatz is now similar to the signal operator of the \ac{isq}-\ac{qsp} multiplied by a signal processor:
\begin{equation}
    \widehat{A_{j + 1}}(\underline{\theta}, \underline{\delta}, \underline{\phi}) = e^{i \phi_{j + 1} \widehat{I} \otimes \widehat{X}} e^{i \delta_{j + 1} \widehat{I} \otimes \widehat{Z}} e^{i \theta_{j + 1} \widehat{H_{i(j)}} \otimes \widehat{Z}} \widehat{A_{j}}(\underline{\theta}, \underline{\delta}, \underline{\phi})
\end{equation}
This ansatz thus has three new parameters to optimize per iteration: $ \{ \theta_{j + 1}, \delta_{j + 1}, \phi_{j + 1} \} $.
This structure allows for generating a quantum circuit ansatz \Cref{circuit_p_adapt_vqe} that can be both similar to \ac{adapt}-\ac{vqe} and to \ac{isq}-\ac{qsp} depending on the value of the optimized parameters.

\subsection{Cost Function on the Post-Selected Subspace}
It is important to underline that the measurement of the Hamiltonian expectation values must be done in the subspace in which the energy is optimized. This allows us to be sure that we target the specific state selected by the projector. The unnormalized expectation value is: 
\begin{equation}
    h = \bra{\psi_{j}} (\ketbra{0} \otimes \widehat{H_{p}}) \ket{\psi_{j}}
\end{equation}

The optimized quantity is the conditional energy of the selected subspace, which is defined via the following Rayleigh quotient:
\begin{equation}
    \bra{\varphi_{j}} \widehat{H_{p}} \ket{\varphi_{j}} 
    = \frac{\bra{\psi_{j}}(\ketbra{0}\otimes\widehat{H_p}\ket{\psi_{j}}}
             {\bra{\psi_{j}}(\ketbra{0}\otimes\widehat{I})\ket{\psi_{j}}},
    = \frac{h}{P_{succ}}
    \label{eq_rayleigh}
\end{equation}

\begin{figure}[tb]
    \begin{center}
        \resizebox{\linewidth}{!}
        {\includegraphics{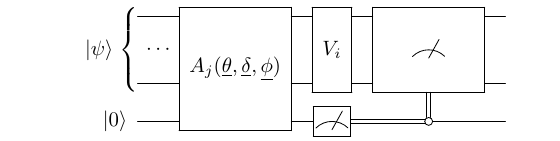}}
    \end{center}
    \caption{Quantum circuit used to measure an expectation value of the state prepared by the Projector-\ac{adapt}-\ac{vqe} ansatz at the $j$th iteration: $ \bra{\varphi_{j}} \widehat{H_{i}} \ket{\varphi_{j}} $.}
    \label{circuit_p_energy_measurement}
\end{figure}

Both $h$ and $P_{succ}$ are estimated from the same measurements, with $P_{succ}$ the fraction of shots accepted by post-selection. A typical measurement circuit is illustrated by \Cref{circuit_p_energy_measurement}.

An intuitive way to grasp why this ansatz may perform better is that it is easier to hide the part of the state vector that contains the high-energy component than to search for a transition from this state to a lower-energy state.

\subsection{Structure of the Ansatz}
In this section, we want to show that this ansatz can be equivalent to either an \ac{adapt}-\ac{vqe} or to an \ac{isq}-\ac{qsp} quantum circuit structure. 

The layer $j$ of the ansatz is: $ \widehat{L_j} = (\widehat{I} \otimes \widehat{S_j}(\phi_{j})) \, \widehat{W_j}(\theta_j, \delta_{j}) $ with $ \widehat{S_j}(\phi_{j}) = \widehat{R_X}(\phi_j) $ acting on the ancilla and
$\widehat{W_j}(\theta_j, \delta_{j}) = (\widehat{I} \otimes\widehat{R_Z}(\delta_j)) e^{i\theta_j\widehat{H_{i_j}}\otimes\widehat{Z}}$ acting on both system and ancilla registers.
Using the spectral decomposition of $\widehat{Z}$, we have 
\begin{equation}
    \widehat{W_j}(\theta_j, \delta_{j}) = \sum_{s\in\{0,1\}} e^{i(-1)^{s} (\theta_j \widehat{H_{i(j)}} + \delta_{j} \widehat{I})} \otimes \ketbra{s}
\end{equation}
This means that the signal operator $\widehat{W_j}$ evolves the system forward on the $\ket{0}$ subspace and backward on $\ket{1}$. The signal processors $\widehat{S_j}$ are the only operations mixing the two subspaces.

We can identify different regimes of our ansatz:
\begin{enumerate}
  \item The ansatz is equivalent to \ac{adapt}-\ac{vqe} if $\forall j, \phi_j=0$. This implies that every signal processor $\widehat{S_j}$ is diagonal, and the ancilla never leaves the state $\ket{0}$.
  \item The ansatz is equivalent to \ac{isq}-\ac{qsp} circuit shown in \Cref{circuit_isq_qsp} when:
  \begin{itemize}
      \item At each layer $j$, the selection of the summands $\widehat{H_{i(j)}}$ exactly matches the ordering of the Trotterization scheme \Cref{eq_trotterization} that gathers the summands in a way that approximate a signal operator.
      \item All the $\delta_{j}$ are equal to the same shift $\delta$.
      \item All the optimized angle $\phi_{j}$ are tuned according to \ac{qsp} phase angle for a boxcard function.
  \end{itemize}
  However, gradient-driven operator selection will not, in general, find it.
\end{enumerate}

\subsection{Success Probability Simplified Estimation}
Post-selection is the price \ac{pva} must pay for its non-unitary filter. This non-unitarity is what allows our ansatz to reach a target accuracy in fewer layers than its unitary baseline.
The post-selection structure of our ansatz is inspired by the one used in fault-tolerant projective algorithms, such as state filtering by \ac{qsp}. 
This structure is composed of a single signal-processing ancilla that is rotated between queries without ever being measured.

This is a structural difference from standard projective ground-state methods that measure and reset an ancilla at every step. \cite{leadbeater_non-unitary_2023, ejima_probabilistic_2025, choi_rodeo_2021}
In those methods, per-step acceptance probabilities multiply, and the total decays as the product of the layer-selection probabilities $p_k$.
Instead, we only need the ancilla qubit to be postselected throughout the circuit.

This subsection describes a model based on reasonable assumptions to justify why sampling is not necessarily an issue for the \ac{pva}.
This non-sampling issue will be verified for specific instances in the experimental section.
Suppose that the variance of the Hamiltonian expectation-value estimator is similar for the two prepared states and for each iteration. 
Then the number of shots $S$ that measure the Hamiltonian expectation value is proportional to the probability of projecting in the good subspace $P_{\mathrm{succ}}$:
\begin{equation}
  \frac{S_{\mathrm{PVA}}}{S_{\mathrm{ADAPT}}} = \frac{1}{P_{\mathrm{succ}}}
  \label{eq_shotratio}
\end{equation}
The total number of shots of the complete ansatz construction increased with the number of ansatz layers $L$.
Assuming that the optimizer requires a similar number of steps to converge within each ansatz layer and that $P_{\mathrm{succ}}$ is stable: 
\begin{equation}
  S \propto L 
\end{equation}
In practice, a fewer-layer \ac{pva} ansatz may require fewer shots for the complete ansatz construction by compensating the higher cost of a layer with a smaller number of layers:
\begin{equation}
  \frac{S_{\mathrm{PVA}}}{S_{\mathrm{ADAPT}}} \times \frac{L_{\mathrm{PVA}}}{L_{\mathrm{ADAPT}}} < 1
\end{equation}
If not, \ac{pva} will have chosen a shorter circuit at the cost of additional measurement.
On most current hardware, this circuit length-sampling speed trade-off can be a good one, since multiqubit gate noise remains the bottleneck.

\section{Experimental Setup \label{sec_expe_set}}

\subsection{Systems and Software}
To evaluate the performance of our \ac{pva} algorithm, we numerically simulated the task of finding the ground-state energy for the following molecules: $H_4$, $LiH$, $H_6$, and $BeH_2$.
All the atoms are placed on a straight line and spaced with the same bond distance.
To have a  strong enough correlation effect, we selected the following bond distances: $ R  =  1.5 \r{A} $ for  $H_4$ and  $BeH_2$;  $ R  =  2 \r{A} $ for $LiH$ and $H_6$. 
No electrons are frozen.
The Hamiltonians were generated using the STO-3G minimal basis set and mapped to qubits with the Jordan-Wigner transformation. 
The number of qubits to map the molecule hamiltonian within this basis are respectiely $8$, $12$, $14$, $12$.
For all simulations, we used the \ac{hf} state as an initial state. Reference energies are obtained by exact diagonalization.

\subsection{Operator Pools}
We benchmarked our \ac{pva} algorithm using two different pools of operators:  the  Qubit-\ac{adapt}-\ac{vqe} pool that uses a pool of individual Pauli strings, and a Fermionic-\ac{adapt}-\ac{vqe} pool that uses fermionic excitation operators, both derived from \ac{hva} operators. The fermionic operators were implemented exactly with the \ac{scb} formalism.
\subsubsection{CNOT by Layer Estimation}
\begin{figure*}[tb]
    \begin{center}
        \resizebox{\linewidth}{!}
        {\includegraphics{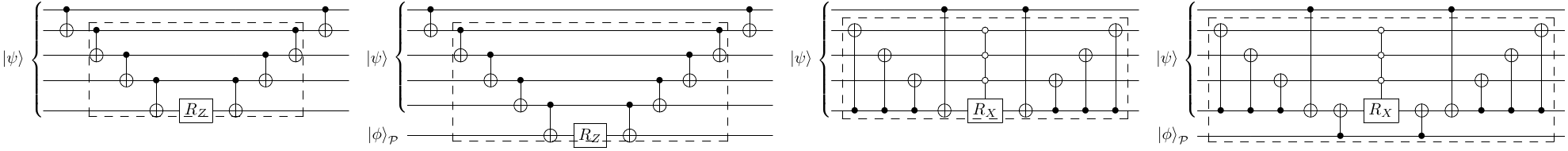}}
    \end{center}
    \caption{Quantum circuit that represents the part (dotted boxes) of the exponentialized operator from which the CNOT count differs ($g$ value) between the different operator pools.
    Respectively from left to right: $ e^{i t \widehat{Z} \widehat{Z} \widehat{Z} \widehat{Z} \widehat{Z}} $, $ e^{i t \widehat{Z} \widehat{Z} \widehat{Z} \widehat{Z} \widehat{Z} \widehat{Z}} $, $ e^{i t (\widehat{Z} \widehat{\sigma} \widehat{\sigma} \widehat{\sigma} \widehat{\sigma} + \widehat{h.c.} )} $, $ e^{i t (\widehat{Z} \widehat{\sigma} \widehat{\sigma} \widehat{\sigma} \widehat{\sigma} \widehat{Z} + \widehat{h.c.})} $.}
    \label{circuit_delta_cx}
\end{figure*}

For all pools used in this paper, the CNOT count can be well approximated by $nCNOT = n \alpha$ with $n$ the number of layers and $\alpha$ the per-layer slope.
This slope is upper-bound by:
\begin{equation}
    \Upsilon_{i} = g_{i} + 2 \times (nb_{om} - 4)
    \label{eq_cnot_count}
\end{equation}
with $nb_{om}$ the number of molecular orbitals.
The difference between the pools comes from the value of the CNOT count $g$:
\begin{itemize}
    \item For original qubit \ac{adapt} pool: $ g_{ZZZZ} = 6 $
    \item For original fermionic \ac{adapt} pool: $ g_{\sigma^{\dag} \sigma^{\dag} \sigma \sigma} = 26 $
    \item For qubit projector \ac{adapt} pool: $ g_{Z ZZZZ} = 8 $
    \item For fermionic projector \ac{adapt} pool: $ g_{Z \sigma^{\dag} \sigma^{\dag} \sigma \sigma} =  28 $
\end{itemize}
These values are computed by transpiling the circuits of \Cref{circuit_delta_cx} and keeping the number of CNOTs.

\section{Results \label{sec_result}}
\begin{figure*}[!p]
    \begin{center}
        \resizebox{!}{\textheight}
        {\includegraphics{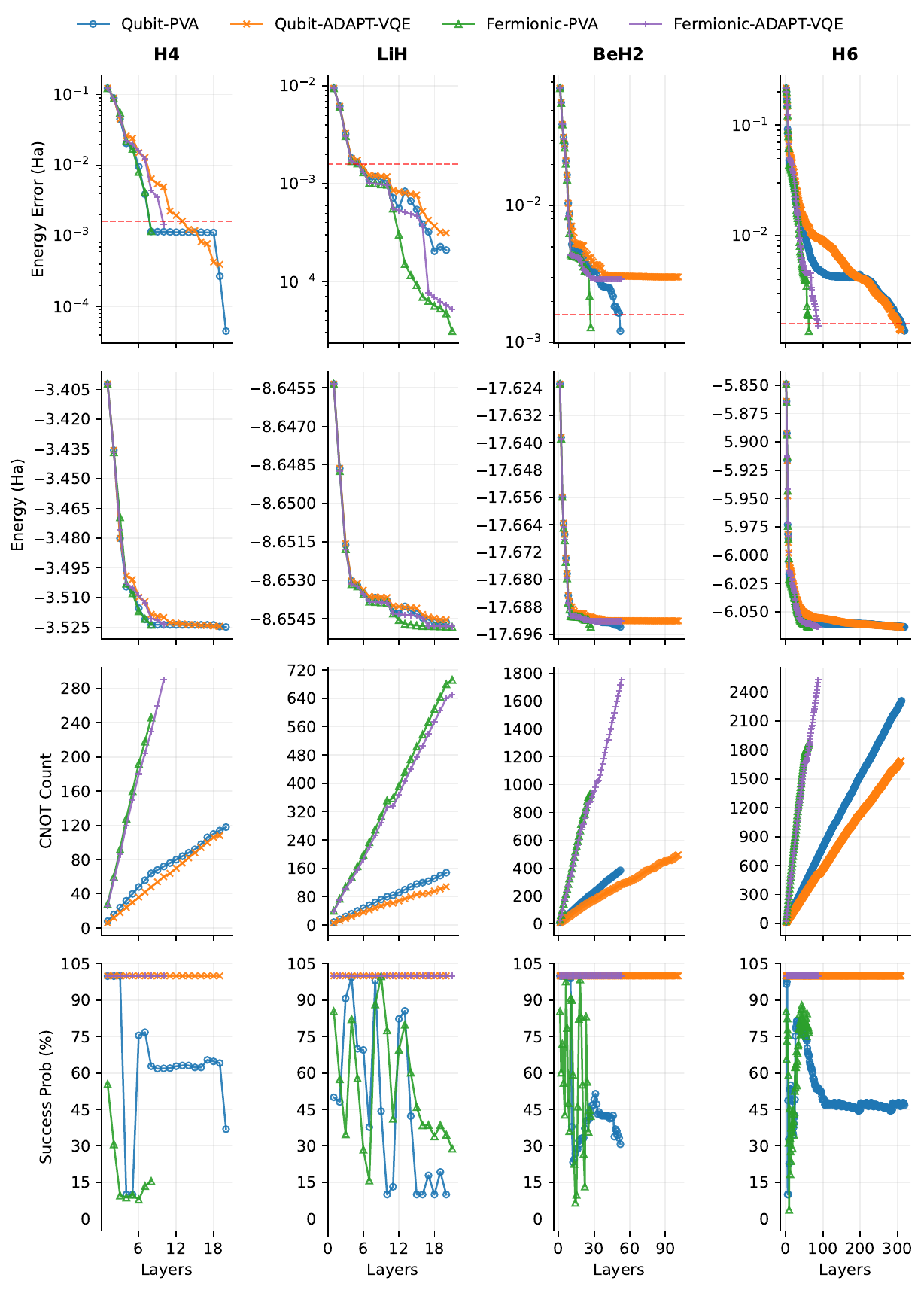}}
    \end{center}
    \caption{Performance comparison of the \ac{pva} versus standard Qubit-\ac{adapt}\ac{vqe} and Fermionic-\ac{adapt}-\ac{vqe}.}
    \label{statevec}
\end{figure*}

We first performed exact statevector simulations using the L-BFGS-B optimizer. In a second time, we used finite sampling and the SPSA optimizer in order to test the viability of our algorithm against statistical noise. In both cases, the analytic gradients and energies were evaluated within the post-selected subspace. \Cref{statevec} compares \ac{pva} with the qubit pool against standard
Qubit-\ac{adapt}-\ac{vqe} and fermionic ac{adapt}-\ac{vqe} under exact statevector simulation.

\subsection{Energy convergence}
The first two rows show the expressivity advantage provided by the non-unitary architecture of our \ac{pva} ansatz. These plots show a monotonic convergence toward the exact FCI ground state on all tested molecules. We observe a reduction in the number of layers required to reach chemical accuracy ($\Delta E < 1.6 \times 10^{-3}$ Ha). For $H_4$, the \ac{pva} converges in $8$ layers compared to the $15$ required by the standard unprojected ansatz (orange), which represents a factor two reduction in required operator depth. For $LiH$, both methods reach chemical accuracy at layer $4$, but \ac{pva} converges faster afterward. The advantage of our method is the most noticeable in the $BeH_2$ system, where the \ac{pva} reaches chemical accuracy in $52$ layers. In contrast, the standard Qubit-ADAPT-VQE becomes trapped in an optimization plateau and struggles to converge even after $200$ layers with the standard \ac{hva} based pool.
Finally, for the highly correlated $H_6$ chain, the \ac{pva} converges roughly twice as fast for the first layers. As it approaches chemical accuracy, the algorithm encounters an optimization plateau and ultimately aligns with the convergence trajectory of the standard Qubit-ADAPT-VQE. This behaviour highlights the rapid early-stage convergence provided by PVA, making our approach advantageous for applications requiring fast, approximate state preparation.
We note that this behaviour illustrates the worst-case performance of our method, which is similar to standard \ac{adapt}-\ac{vqe}.

\paragraph{CNOT Count}
The third row shows the number of controlled-not gates per layer.
Because the \ac{pva} utilizes an ancilla-controlled projection, it introduces an overhead of two CNOT gates per layer compared to the standard Qubit-ADAPT sequence.
This overhead is visible as a slightly steeper slope in the CNOT count.
However, in most cases, the \ac{pva} achieves chemical accuracy in fewer layers, which implies a lower number of total CNOTs.
\Cref{eq_cnot_count} directly explains the ordering between the different curves and why the two-qubit and the two-fermionic slopes have similar behavior.
The most interesting thing that \Cref{eq_cnot_count} teaches us is that the bigger the molecule, the closer to $1$ the ratio between the different curves' slopes:
\begin{equation}
    \lim_{nb_{om} \to\infty} \frac{\Upsilon_{i}}{\Upsilon_{j}} = 1
\end{equation}
with $ \{ i, j \} \in \{ ZZZZ, \sigma^{\dag} \sigma^{\dag} \sigma \sigma, Z ZZZZ, Z \sigma^{\dag} \sigma^{\dag} \sigma \sigma \} $.
It implies that in terms of CNOT gate count, the larger the molecule, the smaller the \ac{pva} strategy overhead becomes with respect to the original \ac{adapt}-\ac{vqe}.
Similarly, the larger the molecule, the smaller the relative CNOT overhead of the fermionic pool is, with respect to the other pools.
It suggests that for molecules larger than our example, the fermionic \ac{pva} is likely to be the most efficient choice in terms of total gate count.
The same kind of reasoning also applies between the operator families that generate the pools.

\paragraph{Subspace Projection Probability}
Finally, the bottom row illustrates the physical mechanics of the subspace projection. As a purely unitary algorithm, standard Qubit-ADAPT-VQE maintains a constant $100\%$ measurement acceptance rate. In contrast, the \ac{pva} applies a filter that makes our success probability ($P_{succ}$) drop.
The plots show that after an initial descent, $P_{succ}$ rapidly stabilizes at tractable operational levels (\textit{e.g.}, stabilizing at $\approx 60\%$ for $H_4$ and $\approx 45\%$ for $H_6$). This empirical data confirms that the \ac{pva} is capable of reducing the number of operators needed without inducing a complete collapse of the subspace projection probability, confirming its viability for \ac{nisq} applications.

\paragraph{Fermionic and Qubit Pools}
We decided to test the robustness of our approach by modifying the pool. By utilizing the fermionic operators instead of the individual qubit transitions, we observe that the convergence advantage is less pronounced for the smaller systems ($H_4$ and $LiH$). However, we still have our plateau-breaking advantage for $BeH_2$, which otherwise struggles to converge with a standard fermionic \ac{hva} pool.
As for $H_{6}$, with the fermionic operators, we are able to maintain our advantage until the end. We reach chemical accuracy in about two-thirds of the iterations needed for standard Fermionic-ADAPT-VQE. Across all cases, we are able to reach chemical accuracy with fewer layers and CNOTs than the canonical version. However, we note that in the fermionic case, our success probability ($P_{succ}$), while still not collapsing entirely, is less stable than with the Qubit pool.

\subsection{Circuit evaluations and Accuracy.}
\begin{figure*}[tb]
    \begin{center}
        \resizebox{\linewidth}{!}
        {\includegraphics{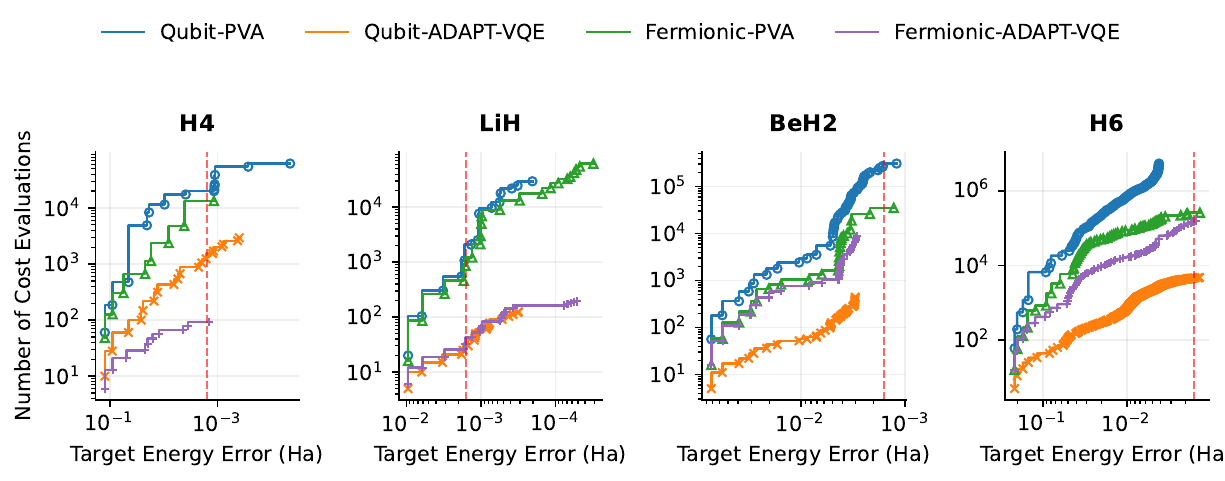}}
\end{center}
\caption{Total number of circuit evaluations with respect to the achieved Energy error.}
\label{graph_evaluations}
\end{figure*}
Using the same statevector trajectories as \Cref{statevec}, \Cref{graph_evaluations} compares the cumulative number of circuit evaluations required to reach a target energy error. 
For each target error, the figure shows all recorded circuit evaluations, including both cost-function optimization and operator-pool screening, up to the first adaptive iteration that reaches the target. 
As \Cref{statevec} experiment is produced using a state-vector emulator, the fair way to compare the number of evaluations is to divide the number of evaluations by the probability to project in the good subspace for each \ac{adapt} iteration $i$: $ nb_{\mathrm{eval}} = \sum_{i} \frac{nb_{\mathrm{eval}, i}}{P_{\mathrm{succ}}} $.
The figure shows that the reductions in final circuit size come at the expense of additional optimization and postselection costs. This relative overhead depends on the problem instance and on the operator pool used.
For the fermionic pool, the relative overhead is less pronounced for $H_6$ and $BeH_2$ than for $H_4$ and $LiH$.
With the exception of $H_6$, the overhead also diminishes for the qubit pool.
These results suggest a more favorable trade-off for the larger instances.
Moreover, for $BeH_2$, the largest system we studied in terms of qubit count, both versions of \ac{pva} overcome the optimization plateau observed for both unitary baselines within the tested optimization budgets. 
This means that the smaller evaluation counts of these baseline runs do not establish a lower cost of reaching the same accuracy.

\subsection{Finite Shots Simulation}

\begin{figure}[tb]
    \begin{center}
        \resizebox{\linewidth}{!}
        {\includegraphics{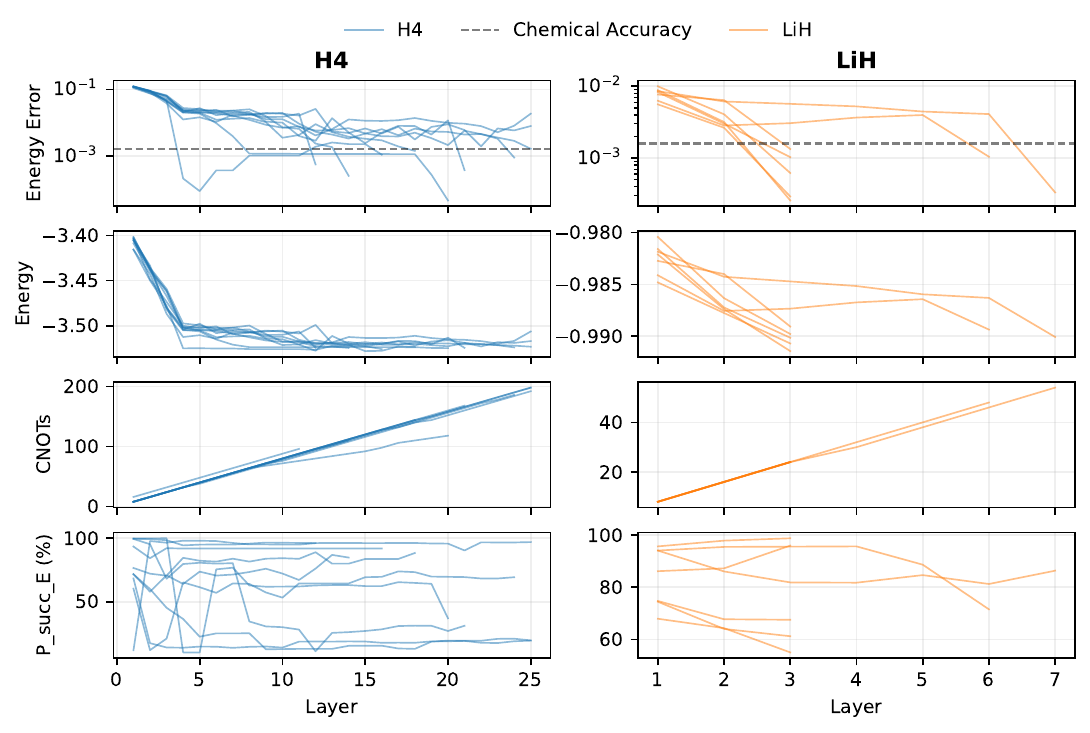}}
    \end{center}
    \caption{Finite shots simulation of our \ac{pva} method for $H_{4}$ and $LiH$}
    \label{shots}
\end{figure}

On \Cref{shots}, we evaluated our \ac{pva} framework using finite measurement shots and the Simultaneous Perturbation Stochastic Approximation (SPSA) optimizer in order to further test the viability of our method. Because of classical simulation overhead, we limited our tests to $H_4$ and frozen core $LiH$. Our SPSA-driven \ac{pva} converged successfully to chemical accuracy in both cases without any artificial cost-function penalties. Even with shot noise, our probability of success stabilizes at a practical level. Our results show the same monotonic convergence as the statevector simulation.
\Cref{shots} shows that, on average, we can keep our gains in the number of layers required. 




\section{Conclusion}
In this paper, we introduce a novel family of ansatzes, the \ac{pva}.
This family of ansatzes combines the operator selection of ADAPT and the signal-processing structure of QSP by using an ancilla-controlled subspace projection.

Our first implementations of the \ac{pva} framework show that we can reach a reduction in the number of operators and CNOTs required to reach chemical accuracy in both Qubit and fermionic pools without increasing the overall CNOT count.
This translates into a reduction of circuit depth. However, this gain in expressivity yields a more complex classical optimization landscape and inherently increases the measurement overhead due to post-selection.
However, our simulations reveal that this does not constitute a practical bottleneck because our success probability does not collapse and often stabilizes at a viable level in our experiments.
By trading measurement time for shallower quantum circuits, the \ac{pva} could constitute a good approach for \ac{nisq} and \ac{isq} architectures.
The evaluation of the impact of physical hardware constraints on our algorithm is planned for future work.

Beyond near-term implementations, the PVA algorithm can serve as an efficient, low-depth state preparation subroutine. Our PVA algorithm could also have applications in quantum machine learning. The nonlinearity of the projection could serve as a quantum analog of classical activation functions in Quantum Neural Networks. Furthermore, the plateau-breaking capabilities and fast initial convergence of our approach on VQAs could be used to mitigate trainability issues and vanishing gradients in deep parameterized quantum circuits.

For future work, we can imagine modifying the ansatz architecture.
Advanced \ac{adapt} \ac{vqe} schemes, such as Tetris or GEO-\ac{adapt}, improve different parts of the scheme. They are not competitors but complementary to \ac{pva}. It is thus possible to combine our \ac{pva} with a more elaborate version of \ac{adapt}.

\bibliographystyle{ieeetr}
\bibliography{biblio}

\section{Fundings}
The work presented in this paper has been supported by AIDAS - AI, Data Analytics and Scalable Simulation - which is a Joint Virtual Laboratory gathering the Forschungszentrum Jülich (FZJ) and the French Alternative Energies and Atomic Energy Commission (CEA). It was also supported by the French PEPR integrated project Etude de la PIle Quantique — EPIQ, (ANR-22-PETQ-0007).

\section{Acknowledgement}
The authors thank Damien BRACONNIER for conducting the experiments presented in Appendix \ref{subappendix_damien}.

\appendix

\subsection{Notations}
The notation used in this paper is gathered in \Cref{tab:pva_symbols}.

\begin{table}[t]
    \caption{Key mathematical symbols and variables used.}
    \centering
    \renewcommand{\arraystretch}{1.3}
    \begin{tabular}{c p{6cm}}
        \hline\hline
        \textbf{Symbol} & \textbf{Definition / Meaning} \\
        $\widehat{H}_p$ & Problem Hamiltonian to be diagonalized \\
        $\widehat{H}_i$ & Summands used to construct the linear combination of Hamiltonian equal to the problem Hamiltonian \\ %
        $\alpha_i$ & Weight coefficients for each summands \\
        $\widehat{V}_i, \widehat{D}$ & Unitary and diagonal matrices such that $\widehat{H}_i = \widehat{V}_i^\dagger \widehat{D} \widehat{V}_i$ \\
        $\lambda_i$ & Eigenvalues of the problem Hamiltonian \\
        $|\lambda_i\rangle$ & Eigenstates associated with $\lambda_i$ \\
        $|\psi\rangle$ & General quantum state vector \\
        $|\psi_i\rangle$ & Initial easily prepared state \\
        $C(\underline{\theta})$ & Variational cost function to be minimized \\
        $\widehat{A}(\underline{\theta})$ & Parametrized unitary matrix (the ansatz) \\
        $\underline{\theta}$ & Vector of variational parameters \\
        $|0\rangle, |1\rangle$ & Ancillary qubit basis states used for flagging solutions \\
        $\widehat{W}_Z$ & Signal operator (query) associated with Hamiltonian simulation \\
        $\gamma$ & Hamiltonian normalization factor ($\gamma = \frac{\pi}{2} \frac{\Gamma}{|\widehat{H_{p}}|}$) \\
        $\delta$ & Shift parameter for eigenvalue projection \\
        $ \delta \pm \frac{\pi}{2} = \gamma \Delta $  & Shift used to move projected eigenvalues \\
        $\mathrm{f}$ & Target boxcar-function approximated by \ac{qsp} \\
        $\widehat{QSP}_Z[\mathrm{f}]$ & Quantum Signal Processing algorithm operator \\
        $\underline{\phi}$ or $\phi_m$ & Phase angles associated with the \ac{qsp} function $\mathrm{f}$ \\
        $m_\phi$ & Number of phase angles used to approximate $\mathrm{f}$ \\
        $\widehat{S}_X(\theta)$ & Signal processor operator ($\widehat{S}_X(\theta) = e^{i\theta\widehat{X}}$) \\
        $\widehat{I}$ & Identity matrix \\
        $\widehat{X}, \widehat{Z}$ & Pauli-X and Pauli-Z matrices \\
        $\widehat{R}_X(2 \theta)$ & Rotation operator around the X-axis \\
        $\underline\theta=(\theta_1, \dots, \theta_L)$ & Underlined symbols denote the vector of all per-layer
    values \\
        \hline\hline
    \end{tabular}
    \label{tab:pva_symbols}
\end{table}

\subsection{Cost-Function Structure}
By linearity, the expectation value of the operator on a quantum state (or the inner product of two states around the operator) can be computed via a weighted sum of all the terms that compose this observable:
\begin{equation}
\begin{aligned}
  \bra{\psi} \widehat{H_{p}} \ket{\varphi} & = \sum_{i} \alpha_{i} \bra{\psi} \widehat{H_{i}} \ket{\varphi} \\
  & = \sum_{i} \beta_{i} \bra{\psi} \widehat{U_{i}} \ket{\varphi}
\end{aligned}
\label{equation_mesure_somme}
\end{equation}

The goal of the following list is to highlight that evaluating \Cref{equation_mesure_somme} is the main piece to construct many \ac{vqa} cost-functions to optimize:
\begin{itemize}
  \item Eigenvalues (\ac{vqe}):
  \begin{equation}
  \begin{aligned}
    C(\underline{\theta}) & = \bra{\psi(\underline{\theta})} \widehat{H_{p}} \ket{\psi(\underline{\theta})} \\
    & = \bra{\psi_{i}} \widehat{A}(\underline{\theta})^{\dag} \widehat{H_{p}} \widehat{A}(\underline{\theta}) \ket{\psi_{i}}
  \end{aligned}
  \end{equation}
  \item Linear System (\ac{vqls}):
  \begin{equation}
  \begin{aligned}
    C(\underline{\theta}) & = || \widehat{H_{p}} \cdot \widehat{A}(\underline{\theta}) \ket{\psi_{b}} - \ket{\psi_{b}} ||^{2} \\
    & = \bra{\psi_{b}} \widehat{A}(\underline{\theta})^{\dag} \cdot \widehat{H_{p}}^{\dag} \cdot \widehat{H_{p}} \cdot \widehat{A}(\underline{\theta}) \ket{\psi_{b}} + \braket{\psi_{b}|\psi_{b}} \\
    & \qquad - \bra{\psi_{b}} \widehat{H_{p}} \cdot \widehat{A}(\underline{\theta}) \ket{\psi_{b}} - \bra{\psi_{b}} \widehat{A}(\underline{\theta})^{\dag} \cdot \widehat{H_{p}}^{\dag} \ket{\psi_{b}} \\
  \end{aligned}
  \end{equation}
  \item State Initialization:
  \begin{equation}
  \begin{aligned}
    C(\underline{\theta}) & = \braket{\psi_{f} | \psi(\underline{\theta})} 
    & = \bra{0} \widehat{U_{\psi}} \widehat{A(\underline{\theta})} \ket{\psi_{i}}
  \end{aligned}
  \end{equation}
  \item Except from a proposal to construct block-encoding and the previously mentioned imaginary time evolution, most of the cost functions are summarized in \cite{cerezo_variational_2021}.
\end{itemize}


\subsection{Ansatz Classification}
\begin{figure}[tb]
\begin{center}
\resizebox{\linewidth}{!}{\includegraphics{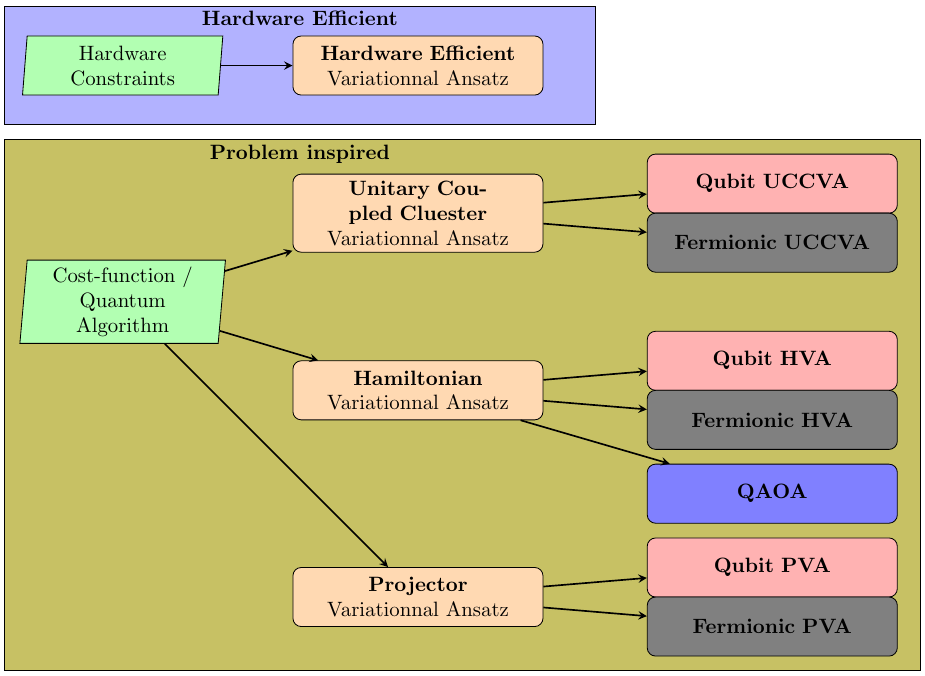}}
\end{center}
\caption{Basic Ansatz Classification Diagram.}
\label{graph_ansatz_zoo}
\end{figure}

The main idea of \Cref{graph_ansatz_zoo} classification and of the following section is to show that the most important ansatz (except \ac{heva}) derives from a truncated and parametrized \ac{isq} quantum algorithm.

\subsection{\acl{heva}}
Hardware-efficient Ansätze have a design that is easily implementable on specific quantum hardware \cite{kandala_hardware-efficient_2017}.
A number of layers $n$ is associated with the expressivity of this ansatz.
Each layer is composed of a sublayer with a two-qubit gate for the entanglement and a layer of single-qubit gates for the state mixing:
\begin{equation}
  \widehat{A}(\underline{\theta}) = \prod_{i}^{n} \widehat{U}(\theta_{i}) \cdot \widehat{U}_{ENT}
\end{equation}
For instance, one can consider the typical example from Qiskit \cite{javadi-abhari_quantum_2024} called 'efficient SU2 ansatz':
\begin{equation}
\begin{aligned}
  \widehat{U}(\theta_{i}) & = \bigotimes_{j = 0}^{nb_{qb} - 1} \widehat{Y_{j}} \widehat{R_{X j}}(\theta_{i, j}) 
  ; \quad
  \widehat{U}_{ENT} & = \prod_{j = 0}^{nb_{qb} - 1} \widehat{CX_{j, j + 1}}
\end{aligned}
\end{equation}
with $nb_{qb}$ the number of qubit.
It can be composed of mostly two-qubit gates acting on physically neighboring qubits.

\subsection{\acl{uccva}}
\acf{uccva} based ansatz are a popular example of physically inspired ansatz \cite{peruzzo_variational_2014, yung_transistor_2014, shen_quantum_2017, mcclean_theory_2016}.
The Unitary Coupled Cluster theory is a chemically inspired method \cite{bartlett_coupled-cluster_2007}.
It can be understood as the Hamiltonian simulation of the allowed electronic transition and so stays in the allowed states:
\begin{equation}
\begin{aligned}
    \widehat{T_{1}}(\underline{\theta_{1}}) & = \sum_{a > b} i \theta_{ab} [ \widehat{a_{a}}^{\dag} \widehat{a_{b}} - \widehat{\mathit{h.c.}} ] \\
    \widehat{T_{2}}(\underline{\theta_{2}}) & = \sum_{a > b > c > d} i \theta_{abcd} [ \widehat{a_{a}}^{\dag} \widehat{a_{b}}^{\dag} \widehat{a_{c}} \widehat{a_{d}} - \widehat{\mathit{h.c.}} ] \\
  \widehat{T}(\underline{\theta}) & = \sum_{i = 0}^{k} \widehat{T_{i}}(\underline{\theta}) ; \qquad
  \widehat{UCC}(\underline{\theta}) = e^{i \widehat{T}(\underline{\theta})}
\end{aligned}
\end{equation}
with $\widehat{a}$ the fermionic ladder operator, $ \underline{\theta} = \underline{\theta_{1}}|\underline{\theta{2}} $ and $\widehat{T_{i}}(\underline{\theta_{i}})$ the operator associated with $i$ electron simultaneous transition. 
Note that we use a specific notation convention for the $\widehat{T}_{i}$ operators so that they are hermitian and that at the end it is written the Hamiltonian simulation of a hermitian matrix.
When these operators are used out of quantum computing domain, the usual notation uses no complexe number and the exponential of a skew-hermitian is taken.
After a Trotterization of order $m$, it leads to the following ansatz:
\begin{equation}
  \widehat{A}_{UCC}(\underline{\theta}) = \prod_{j = 0}^{m} \prod_{i = 0}^{k} e^{i \widehat{T_{i}}(\underline{\theta}_{i, j})}
\end{equation}
Generally, it is specified UCC-S, D, and T, which means there is a notation to the: Singles, Doubles, or Triples $\dots$ simultaneous electronic transition referred to as the cluster order $k$.

More hardware-adapted variants of \ac{uccva} based ansatz exist \cite{babbush_low-depth_2018}.
Often, physically inspired Ansätze are based on symmetry-preserving properties to explore a reduced space \cite{barkoutsos_quantum_2018, gard_efficient_2020, herasymenko_diagrammatic_2021}. 
Controlled expansion of this parameter space is also possible, such as Optimal-Control-inspired Ansatz \cite{choquette_quantum-optimal-control-inspired_2021}, which contains symmetry-breaking unitaries introduced in other physically inspired ansatz. 

\subsection{Digitalized \acl{aqc} based Ansatz}
This subsection describes the path that leads from \ac{aqc} to the \ac{qaoa} original ansatz, as well as the \ac{hva} ansatz \cite{farhi_quantum_2000, mcclean_theory_2016}.
In both cases, the main idea for constructing these ansätze is to truncate the free (or convergence \cite{ronfaut_numerical_2026}) parameters and optimize them.

\textbf{\acl{aqc}} is a procedure done by quantum annealers to go from the known ground state of a chosen quantum system $\ket{\psi_{i}}$ (mixing Hamiltonian: $\widehat{H_{M}}$) toward the unknown ground state of a target quantum system $\ket{\psi_{f}}$ (problem Hamiltonian: $\widehat{H_{p}}$).
This second quantum system can encode an optimization problem or another more complex quantum system.
The adiabatic theorem guarantees, under some conditions, that a system stays in its ground state if the Hamiltonian is transformed slowly enough:
\begin{equation}
\begin{aligned}
  & \ket{\psi_{f}} = \mathcal{T}[e^{- i \int_{t = 0}^{t_{f}} \widehat{H}(t) dt}] \ket{\psi_{i}} \\
  \text{ with: } & \widehat{H}(t) = (1 - r(t)) \widehat{H_{M}} + r(t) \widehat{H_{p}} \\
\end{aligned}
\end{equation}
with $ r(0) = 0 $, $\mathcal{T}$ the time ordering operator, and $ r(t_{f}) = 1$.
$r(t)$ is continuous and evolves slowly enough.
Slowly enough is defined in terms of the spectral properties of $ \widehat{H}(r) $.

\textbf{Digitalized \acl{aqc}} is the same algorithm but discretized with two parameters in order to work on a gate-based quantum computer:
\begin{equation}
\begin{aligned}
  \mathcal{T}[e^{- i \int_{0}^{t_{f}} \widehat{H}(t) dt}]
  & \simeq e^{- i \sum_{m = 0}^{s} \widehat{H}(m \times \delta_{t})} 
  & \sim \prod_{m = 0}^{s} e^{- i \widehat{H}(m \times \delta_{t})}
\end{aligned}
\end{equation}
with the two discretization parameters: $ s = t_{f} / \delta_{t} $ the number of time steps and $n$ the trotter number:
\begin{equation}
\begin{aligned}
  e^{- i \widehat{H}(r)} 
  & = e^{- i ((1 - r) \widehat{H_{M}} + r \widehat{H_{p}})} 
  & \simeq \left( e^{- i \frac{1 - r}{n} \widehat{H_{M}}} \cdot e^{- i \frac{r}{n} \widehat{H_{p}}} \right)^{n}
\end{aligned}
\end{equation}
Trotterization also discretizes the problem and the mixing Hamiltonians if they cannot be implemented exactly.

When transformed into a quantum ansatz, all the arbitrary rotations become classically optimized parameters:
\begin{equation}
\begin{aligned}
  \widehat{AQC} & = \prod_{m = 0}^{s} \left( e^{- i \frac{1 - r(m \times \delta_{t})}{n} \widehat{H_{M}}} \cdot e^{- i \frac{r(m \times \delta_{t})}{n} \widehat{H_{p}}} \right)^{n} \\
   \longrightarrow
   \widehat{A}_{AQC}(\underline{\gamma}, \underline{\beta}) & = \prod_{m = 0}^{s} \left( e^{- i \gamma_{m} \widehat{H_{M}}} \cdot e^{- i \beta_{m} \widehat{H_{p}}} \right)
\end{aligned}
\end{equation}
with $ \gamma_{m} $ and $ \beta_{m} $ the classically optimized free parameters.

\subsubsection{\acl{qaoa}}
\ac{qaoa} is a specific case for binary optimization problems.
The original proposal shows instances for which it behaves well \cite{farhi_quantum_2014} and \cite{farhi_quantum_2019} provided complexity-theoretic explanations of the hardness of modeling this algorithm classically.
It uses the mixing Hamiltonian $ \widehat{H_{M}} = \sum_{i = 0}^{nb_{qb} - 1} \widehat{X}_{i} $
and $\widehat{H_{p}}$ a diagonal Hamiltonian, often corresponding to an Ising problem.
These two Hamiltonians can be implemented exactly.
\ac{qaoa} is one of the most studied \ac{vqa}. \cite[Table 1]{blekos_review_2024} provides a summary of the main variants of its associated ansatz.
It is possible to encode hard constraints using, for instance, Quantum Alternating Operator Ansatz \cite{hadfield_quantum_2019} or Fermionic quantum approximate optimization algorithm \cite{yoshioka_fermionic_2023}.

\subsubsection{\acl{hva}}
The \ac{hva} is the generalization of the previous technique to arbitrary Hamiltonians: initially designed for quantum systems (Fermi-Hubbard) diagonalization \cite{wecker_progress_2015}, associate resource estimation \cite{cade_strategies_2020}, impact of errors \cite{reiner_finding_2019}, optimization guarantee \cite{park_hamiltonian_2024}, more general problems \cite{wiersema_exploring_2020}, and quantum state preparation \cite{ho_efficient_2019}.
When applied to a chemical problem, variational algorithms generally start from the Hartree-Fock state, which is the solution of the final Hamiltonian without all the electron-electron interactions diagonalized. 
The mixing (and initial) Hamiltonian is thus the problem Hamiltonian without the electron-electron terms, more generally, only the matrix diagonal of the final Hamiltonian.
The initial Hamiltonian is thus included in the final one.
When the ansatz is developed, it leads to the Hamiltonian simulation of all the terms that compose the initial Hamiltonian:
\begin{equation}
  \widehat{H_{p}} = \sum_{i = 0}^{\mathbf{S} - 1} \gamma_{i} \widehat{H}_{i} \longrightarrow \widehat{A}_{HVA}(\underline{\theta}) = \prod_{r = 1}^{s} \prod_{i = 0}^{\mathbf{S} - 1} e^{- i \theta_{i, s, r} \widehat{H_{i}}}
\end{equation}
with $ \theta_{i} $ the classically optimized free parameters.
Note that due to Trotter error, the order in which the exponentiated terms are combined impacts the efficiency of this ansatz.
It also has some specific variants, such as Number-preserving and Efficient \ac{hva} \cite{cade_strategies_2020}.

\subsection{Extra Graphs: Two-Parameters per Layer Ansatz \label{subappendix_damien}}
\begin{figure}[tb]
\begin{center}
\resizebox{\linewidth}{!}{\includegraphics{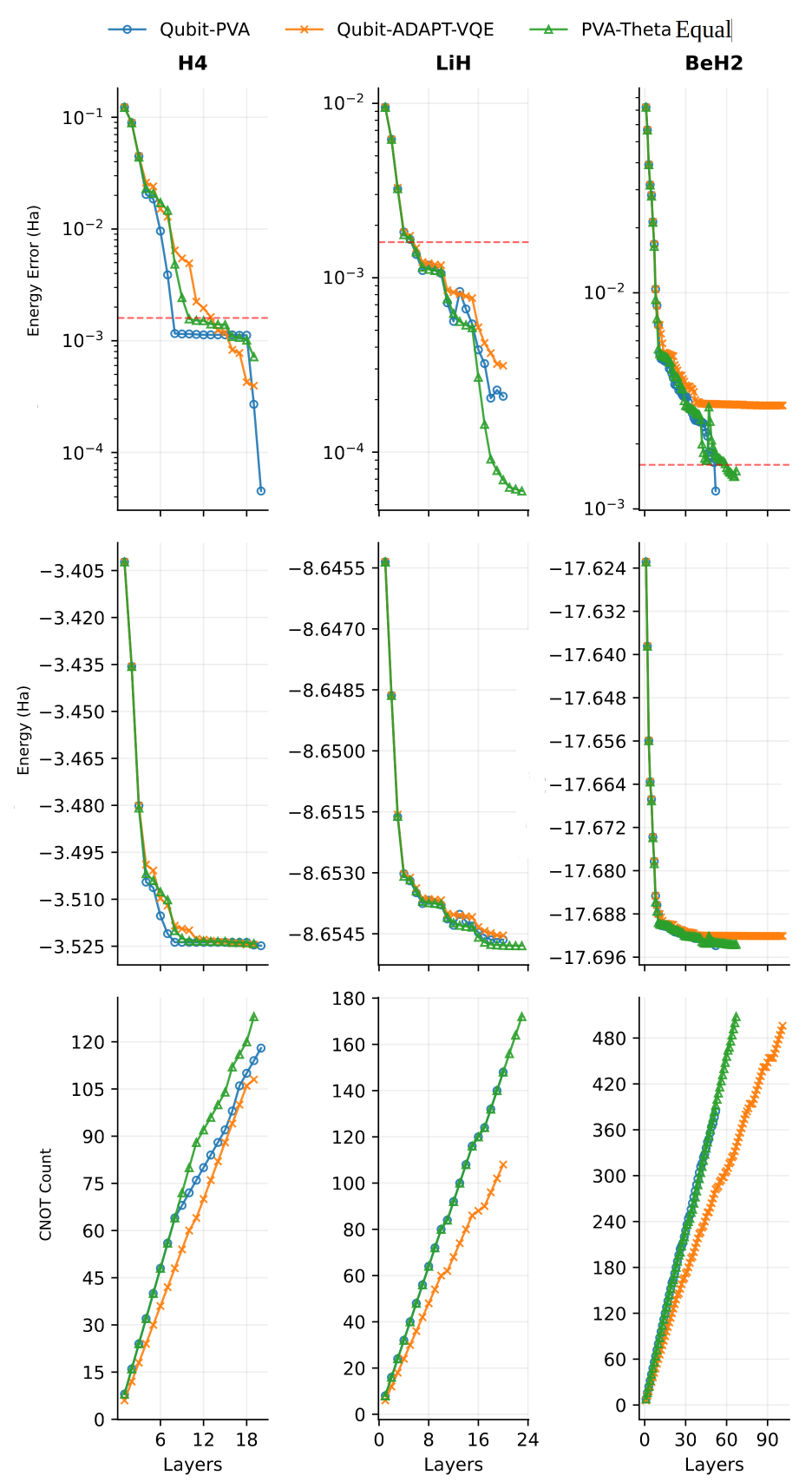}}
\end{center}
\caption{Performance comparison of the Qubit-\ac{pva} versus $2$-parameters-Qubit-\ac{pva}, and versus standard Qubit-\ac{adapt}\ac{vqe}.}
\label{graph_2parametre}
\end{figure}

This sub-appendix presents extra results obtained by testing a variant of the qubit-\ac{pva} for which there are only two free parameters, as the value of $\delta_{i}$ and $\theta_{i}$ are equal for all any ansatz layer $i$ in \Cref{graph_2parametre}.
These results show that the number of parameters per layer of the \ac{pva} can be reduced while still providing a circuit-depth advantage.

\subsection{Implementation Details}
\subsubsection{System Generation and Software Stack}
To compute all electronic structure quantities and the Hartree-Fock method, we used the classical PySCF driver. We implemented the quantum circuit construction and statevector simulations using the Qiskit and Qiskit Nature software stacks. For the finite-shot executions, we used Qiskit's V2 Primitives (SamplerV2) and the Aer simulator. The transpilation was done using Qiskit's transpile function with optimization level $1$ and [cx, rz, sx, x] as basis gates. All the molecular Hamiltonians were generated using the STO-3G minimal basis set. We mapped the active space to qubits via the Jordan-Wigner transformation. For all simulations, the Hartree-Fock state was chosen as the initial reference state. The energy error $\Delta E$ was computed using exact classical diagonalization of the Hamiltonians.

\subsubsection{Operator Pools}
The adaptive pools were derived from \ac{hva} excitation operators. For the Fermionic-ADAPT-VQE, we kept all excitations as an Hamiltonian simulated operator. For the Qubit-ADAPT-VQE,  we decomposed the fermionic operators to extract individual Pauli strings. 
To optimize the pool size, we discarded purely diagonal $Z$-strings.
Concerning the Fermionic Pool, the quantum circuit used to implement the fermionic operator is constructed using the \ac{scb} formalism introduced by \cite{ollive_gate_2025}.
The authors want to highlight that this construction strategy differs from the standard implementations of fermionic operators using Pauli evolution. In our work, fermionic transitions are implemented exactly, so there is no possibility of leaving the solution space. 

\subsubsection{Optimisation and Measurement Strategies}
For the exact-statevector simulations, the parameter optimization ($\theta, \phi, \delta$) was performed globally at each iteration using the L-BFGS-B algorithm with a gradient tolerance of $10^{-5}$. 
We used the chemical accuracy as a stopping criterion for large molecules. 
We let our simulations run for at least $20$ iterations and if the optimisation gets stuck on a barren plateau, we let it ran for $200$ iteration before stopping.
For the shot simulation, we used the SPSA optimizer. The SPSA algorithm was configured with automatic calibration of the learning rate and perturbation step size, and with a maximum of $200$ iterations per layer. To minimize the quantum measurement overhead in these finite-shot simulations, we used Pauli grouping techniques to enable the simultaneous evaluation of commuting observables.
Expectation values for the energy and the gradient approximations were evaluated using a total finite sampling budget of $5\times10^5$ measurement shots per circuit execution including postselection.



\end{document}